\pdfoutput=1
\documentclass[letterpaper,10pt]{article} 

\usepackage{opticameet3} 

\usepackage{cite}
\usepackage{amsmath,amssymb,amsfonts}
\usepackage{algorithmic}
\usepackage{graphicx}
\usepackage{textcomp}
\usepackage{xcolor}
\usepackage[linesnumbered,ruled,vlined]{algorithm2e}
\usepackage{svg}
\usepackage{longtable}
\usepackage{float}
\usepackage{caption}
\usepackage{subcaption}
\usepackage{multirow}
\usepackage[export]{adjustbox}
\usepackage{soul}

\SetCommentSty{mycommfont}

\SetKwInput{KwInput}{Input}                
\SetKwInput{KwOutput}{Output}              
\SetKwFunction{KwFn}{Fn}                   
\usepackage[colorlinks=true,bookmarks=false,citecolor=blue,urlcolor=blue]{hyperref} 

\svgpath{./images/}

\begin{document}


\title{Low-Latency Upstream Scheduling in Multi-Tenant, SLA Compliant TWDM PON}

\vspace{-6mm}
\author{Arijeet Ganguli and Marco Ruffini}
\vspace{-1mm}
\address{School of Computer Science and Statistics, Trinity College Dublin, Ireland}
\vspace{-1mm}
\email{gangulia@tcd.ie, marco.ruffini@tcd.ie} 

\vspace{-6mm}

\begin{abstract}
We present a multi-tenant multi-wavelength upstream transmission scheme for virtualised PONs, enabling compliance with latency-oriented Service Level Agreements (SLAs). Our analysis highlights an important trade-off between single-channel vs. multi-channel PONs, depending on ONUs tuning time. 
\end{abstract}

\vspace{-2mm}
\section{Introduction}
\vspace{-2mm}

Over the past decade, network architectures have transitioned from closed systems to open and disaggregated systems. Software Defined Networking (SDN) and Network Function Virtualization (NFV) have played a key role, offering programmability, cost efficiency and flexibility. Passive Optical Networks (PONs) have experienced a similar transition towards virtualisation \cite{MEC,FASA}, which enables dynamic software-based control of scheduling algorithms and multi-tenancy. This means independent Virtual Network Operators (VNOs) are able to run multiple independent upstream scheduling algorithms in a shared PON.

In order to support 5G and beyond use cases and business applications (including fronthaul for different type of mobile services), which require low latency guarantee, PONs will need to support Service Level Agreements (SLAs). This means that capacity sharing cannot be achieved at the expense of latency performance. This issue has been addressed by work in \cite{FASA}, where DBA algorithms can be modified to suit specific applications, and in \cite{vdba}, with the development of a virtual Dynamic Bandwidth Allocation (vDBA) architecture, which allows tenants to specify indiviual allocation grant with microsecond precision.

In this work we extend our vDBA concept to a multi-channel PON network, where the merging of Bandwidth Maps (BMaps) is carried out across multiple wavelength channels. The OLT, besides indicating the time at which the ONUs can transmit, it also indicates the wavelength channel. 
The key objective is to define a number of SLAs, expressed in terms of maximum latency and compliance level, and to propose an algorithm that can operate the Bandwidth Map merging operation, minimising the probability to breach SLAs, across all VNOs.

The evaluation of our work is based on two key performance analysis. 
First we compare the latency for systems using different line rates and number of channels. Assuming an overall PON capacity of 200Gb/s, we consider a system with 8 channels at 25Gb/s, one with 4 channels running at 50Gb/s, and one with a single channel at 200Gb/s (being single channel there is no multi-wavelength allocation for the 200Gb/s option).
Then we show how the difference in performance changes when we consider different tuning times for the ONU transmitters. For this we compare the case of negligible tuning time (i.e., if the system implements channel bonding, \cite{channel_bonding}, so that the ONU has more than one transceiver always ready to transmit at least at another wavelength); the case of Class 1 transmitters (i.e., tuning time $<10 \mu s$) \cite{NGPON2}, with tuning time of 250 ns (i.e., close to the burst overhead time) and 1 us; and a Class 2 transmitters (i.e., tuning time $>10 \mu s$ but $<25 ms$), with tuning time of 15 us. Class 3 devices are not considered has they have tuning times longer than 25 ms, which is more than two order of magnitudes higher than the PON frame, and would in practice represent a semi-static allocation.

Our results show that our proposed multi-wavelength algorithm is capable of maintaining high percentage of SLA compliance, even for high network load. 
In addition, we highlight an important trade-off between the latency of multi-channel systems and the transmitters' tuning time. If the tuning time is negligible compared to the burst overhead, the multi-channel system has better latency performance. This is due to the fact that the burst overhead tends to have similar time duration independently of the line rate \cite{50G}, thus higher line rate channels require a proportionally larger amount of bits than lower line rate channels. However as the ONU tuning time increases, the single channel system outperforms the multi-channel ones.

\vspace{-3mm}
\section{System architecture}  
\vspace{-2mm}

The use case and architecture addressed in this work is shown in Fig. \ref{fig:twdm}. In our approach, multiple VNOs run different schedulers in parallel (vDBAs), each forwarding a virtual Bandwidth Map (vBMap), which allocates upstream transmission slots to a group of ONUs. This for example allows multiple mobile operators to run fronthaul services (i.e., 7.2 RAN split), using independent Cooperative Transport Interfaces (CTI) 
The scheduling hypervisor (or merging engine) collects all such virtual bandwidth maps to create a single physical bandwidth map, resolving any collisions between slots overlapping in time. Such collisions can be resolved by delaying some of the grants (at the risk of breaching SLAs). In this multi-wavelength approach, the OLT also has the option to select transmission over different channels, to minimise grant delay, although this is constrained by the ONU tuning time. 
Our merging engine makes these decisions based on specific Service Level Agreements (SLAs), so that it minimises the probability of breaching SLAs (which is key for supporting 5G and future 6G services). The use of a stateful algorithm, which takes into consideration the history of a service flow when making scheduling prioritisation decisions, is preferred to stateless algorithms\cite{stateful}. This is because a stateful algorithm can prioritise flows depending on how close they are to breaching their specific SLA target \cite{sla_hypervisor}. 

Thus in this work, we propose a heuristic stateful TWDM scheduling algorithm. 
The objective of the algorithm is to maximise the SLA compliance across all the flows during upstream transmission. We focus specifically on the additional latency introduced by the multi-sharing aspect of the PON. An SLA breach occurs when a given flow accumulates a number of delayed upstream slots that is above its target SLA threshold. For example for an SLA with maximum merging delay of 25 $\mu s$ with 99\% compliance, every time an upstream slot is delayed by more than 25 $\mu s$ with respect to the requested time slot in the virtual BMap, we increment a counter. If the counter goes above the non-compliance rate (in this case 100-99=1\%), calculated over a number of frames (i.e., we use a 1ms window, which is the time duration of a 5G sub-frame), then we consider that an SLA breach has occurred. 

The problem can be formulated as Mixed Integer Programming (MIP) as shown in Fig. \ref{fig:milp}. Equation (1) calculates the maximum delay of any given allocation to remain within the target threshold for SLA breach. We maintain a 4 dimensional binary decision variable matrix \textbf{\textit{X}} where the objective function is explained as follows - the inner truth value function calculates the packet level breach and the outer truth value function calculates the flow level breach across each virtual BMap. Equation (3) is the constraint that any particular channel at any time slot transmits at most 1 BMap allocation. Equation (4) is the constraint that all BMap allocations are alloted unique channel time-slot pairs. Equation (5) checks the conservation of BMap allocations within each virtual BMap.      

\begin{figure}
\hspace{-6mm}
\begin{minipage}[]{0.43\textwidth}
\vspace{-8mm}
\centering\includesvg[width=75mm,height=75mm]{OLT_diag}
\caption{TWDM multi-tenant upstream scheduling algorithm}
\label{fig:twdm}
\end{minipage}
\hspace{4mm}
\begin{minipage}{0.57\textwidth}
\vspace{-8mm}
\centering\includesvg[width=95mm, height=9cm]{MILP_formulation}
\vspace{-6mm}\hspace{2cm}
\caption{MILP Notations and Equations}
\label{fig:milp}
\end{minipage}
\vspace{-8mm}
\end{figure}

The algorithm maintains a few key data structures. A flow-breach likelihood table keeps track of how far each traffic flow is from breaching its SLA (i.e., going above non-compliance rate); this is important to minimise SLA breach, as the algorithm will prioritise scheduling for the flows that are closer to breach their SLA. A channel freetime table maintains and updates the earliest free time of each channel over time, i.e., when the channel can be reallocated to a different BMap allocation. We also keep track of the earliest free time of the various transceivers of the ONUs (i.e., depending on the ONU tuning time). 


With reference to Fig. \ref{fig:pseudocode}, the algorithm first calculates the allocation maxtime which is the latest time an allocation can be scheduled within its latency target (code lines 1 to 3). Then, it starts allocating slots to the various allocations according to the time assigned by their originating virtual BMaps (code lines 5 to 6). Next, it resolves collisions (lines 8 to 21) by allocating slots first in increasing order of non-compliance rate (line 18), then increasing order of their maxtimes (line 19) and finally increasing order of their sizes (line 20). Lines 22 to 43 traverse the sorted BMap and allocates channel and scheduling time for the allocations. Lines 23 to 26 first get the allocation ONU id, then use it to get the free times of all the transceivers of the respective ONU, then allocates the earliest free transceiver and receiver for transmission of the BMap allocation. Line 27 checks the channel free time table and allocates the earliest free wavelength for transmission. Lines 29 to 34 check if tuning of the transceiver to a new channel is required. Lines 35 and 36 calculate the earliest transceiver and receiver free times, respectively. Lines 37 to 40 calculate the scheduling time for transmission according to whether tuning is required or not and lines 41 to 43 assign the transceiver id, receiver id and scheduling time memory variables of the BMap allocation object. Finally, the lines 44 to 47 recalculate the non-compliance rate of all the flows and updates the flow-breach table for scheduling of the allocations in the next time frame.

\vspace{-3mm}
\section{Performance Evaluation}
\vspace{-2mm}
In order to set up the simulation environment, we generate input BMaps from different VNOs. The allocation load on the shared PON was considered for 20\%, 50\% and 80\% of the total upstream capacity (i.e., 200Gb/s across all channels
considered). For each of this allocation loads, we then varied the percentage allocated to SLA-driven flows from 10\% to 100\% of the total load (the remaining part is allocated to best effort flows). We consider 5 VNOs (each one generating a virtual BMap every frame (125 $\mu s$)) and 2 types of SLAs: one requiring 90\% compliance with a latency target of 12.5 $\mu s$ and one requiring 95\% compliance with a latency target of 25 $\mu s$. Each bandwidth map has a uniformly distributed set of allocations, with average burst size 
set at 6\% of the total frame size for a 25Gb/s line rate (the same ONU is also allowed to provide multiple burst per frame). An empty time slot of 0.33 $\mu s$, is introduced between allocations to account for guard time between upstream transmissions. Each ONU has one transceiver, which is tunable on any of the available channels, and constrained by a tuning time, which is a simulation parameter. As mentioned above, the system under investigations, are 8 x 25Gb/s, 4 x 50Gb/s and 1 x 200Gb/s channels. The tuning times considered span from negligible to 250 ns, 1 $\mu s$ and 15 $\mu s$.

The experiment was run for 5000 time frames and the average number of SLA breaches is recorded. 
The results show the ability of the system to comply with the SLAs (y axis) versus the percentage of total flows that require SLA. The different plots in the same figure consider PON loads of 20\%, 50\% and 80\% of the total capacity.

The plot in Fig. \ref{fig:plot_0us} reports the ability to meet SLA for negligible channel tuning time. 
We can see that up to 50\% load there is always full compliance (except a slight drop for SLA flows above 90\%). For 80\% load, compliance starts dropping from 60\% of SLA traffic for the 8X25G and 4x50G systems and from 50\% for the 1x200G. As mentioned above the multi-channel approaches outperform the single channel as at 200G rate the burst overhead is proportionally larger (i.e., in terms of bits required).
As the tuning time increases, in Fig. \ref{fig:plot_250ns}, \ref{fig:plot_1us} and \ref{fig:plot_15us}, we see that the latency performance of the multi-channel system decreases, as the algorithm progressively looses the ability to avoid scheduling delays by tuning on different channels.

\vspace{-2mm}
\begin{figure}[]
\begin{minipage}{1\textwidth}
\vspace{-20mm}
\begin{minipage}{0.6\textwidth}
\vspace{0mm}
\centering\includesvg[width=97mm,scale=0.7]{pseudocode}
\vspace{-7mm}
\caption{Pseudocode for TWDM}
\label{fig:pseudocode}
\end{minipage}
\vspace{-60mm}\hspace{0mm}
\begin{minipage}{0.5\textwidth}
\vspace{-8mm}\hspace{-9mm}
\centering\includesvg[width=60mm, height=53mm]{plot_0us}
\vspace{-5mm}\hspace{-3mm}
\caption{Tuning time: 0s}
\label{fig:plot_0us}
\vspace{-5mm}
\end{minipage}
\end{minipage}
\vspace{-23mm}
\begin{minipage}{1\textwidth}
\vspace{25mm}\hspace{-1mm}
\begin{minipage}{0.3\textwidth}
\vspace{20mm}\hspace{-13mm}
\centering\includesvg[width=60mm, height=45mm]{plot_250ns}
\vspace{-3mm}\hspace{-5mm}
\caption{Tuning time: 250ns}
\label{fig:plot_250ns}
\end{minipage}
\vspace{33mm}\hspace{5mm}
\begin{minipage}{0.3\textwidth}
\vspace{20mm}\hspace{55mm}
\centering\includesvg[width=60mm, height=45mm]{plot_1us}
\vspace{-7mm}\hspace{15mm}
\captionsetup{margin={10mm,0mm}}
\caption{Tuning time: 1$\mu s$}
\label{fig:plot_1us}
\end{minipage}
\vspace{31mm}\hspace{9mm}
\begin{minipage}{0.3\textwidth}
\vspace{20mm}\hspace{40mm}
\centering\includesvg[width=60mm, height=45mm]{plot_15us}
\vspace{-7mm}\hspace{8mm}
\captionsetup{margin={10mm,0mm}}
\caption{Tuning time: 15$\mu s$}
\label{fig:plot_15us}
\end{minipage}
\end{minipage}
\vspace{-50mm}
\end{figure}
\vspace{-1mm}

\section{Conclusions}
\vspace{-2mm}
In this work, we presented a heuristic stateful  algorithm for upstream scheduling in a TWDM multi-tenant PON, capable of satisfying SLAs. We show the role that ONU tuning time plays in terms of latency performance in the trade-off between higher capacity single channel and lower capacity multi-channel systems. While we believe that there is room for improvement in multi-channel scheduling algorithms for longer tuning times, the current results show that tuning times between 250 ns and 1 $\mu s$ would be required to maintain satisfactory performance in multi-tenant, multi-channel PON systems.

{
\fontdimen2\font=2.0pt
\noindent\textbf{Acknowledgments.}  Support from SFI grants 12/RC/2276\_p2, 17/CDA/4760 and 13/RC/2077\_p2 is acknowledged.
}
\vspace{-8mm}


\end{document}